\pdfoutput=1

\documentclass[11pt,a4paper]{article}
\usepackage{jheppub}
\usepackage{slashed}
\usepackage{amsmath}
\usepackage{amssymb}
\usepackage{epsfig}
\usepackage{feynman}

\usepackage{graphicx}
\usepackage{subfigure}
\input{epsf}
\usepackage{epstopdf}

\newcommand{\labell} [1] {\label{#1}}

\def\({\left(} 
\def\){\right)}
\def\[{\left[} 
\def\]{\right]}
\newcommand{\non}{\nonumber \\}

\newcommand{\ie}{{\it i.e.,}\ }
\newcommand{\eg}{{\it e.g.,}\ }

\newcommand{\be}{\begin{equation}}
\newcommand{\ee}{\end{equation}}
\newcommand{\bea}{\begin{eqnarray}}
\newcommand{\eea}{\end{eqnarray}}

\newcommand{\mt}[1]{\textrm{\tiny #1}}

\newcommand{\bt}{\beta}

\catcode`\@=12

\newcommand{\al}{\alpha}
\newcommand\x{{\bf x}}
\newcommand\y{{\bf y}}

\def\del          {\partial}

\newcommand{\reef}[1]{(\ref{#1})}
\renewcommand{\eqref}[1]{(\ref{#1})}

\def\ph1{\phantom{1}}

\newcommand{\beq}{\begin{equation}}
\newcommand{\eeq}{\end{equation}}
\newcommand{\ba}{\begin{aligned}}
\newcommand{\ea}{\end{aligned}}
\newcommand{\beqa}{\begin{eqnarray}}
\newcommand{\eeqa}{\end{eqnarray}}
\newcommand{\beqar}{\begin{eqnarray*}}
\newcommand{\eeqar}{\end{eqnarray*}}

\usepackage{comment}

\title{Entanglement and RG in the $O(N)$ vector model}

\author[a]{Chris Akers,}
\author[b]{Omer Ben-Ami,}
\author[c]{Vladimir Rosenhaus,}
\author[a]{Michael Smolkin,}
\author[b]{and Shimon Yankielowicz}

\affiliation[a]{Center for Theoretical Physics and Department of Physics,\\
University of California, Berkeley, CA 94720 }

\affiliation[b]{Raymond and Beverly Sackler Faculty of Exact Sciences, School of Physics and Astronomy, Tel- Aviv University, Ramat-Aviv 69978, Israel}

\affiliation[c]{Kavli Institute for Theoretical Physics, \\
 University of California, Santa Barbara, CA 93106}

\emailAdd{cakers@berkeley.edu}
\emailAdd{omerben@post.tau.ac.il}
\emailAdd{vladr@kitp.ucsb.edu}
\emailAdd{smolkinm@berkeley.edu}
\emailAdd{shimonya@post.tau.ac.il}

\preprint{TAUP-3003/15}

\abstract{
We consider the large $N$ interacting vector $O(N)$ model on a sphere in $4-\epsilon$ Euclidean dimensions. The Gaussian theory in the UV is taken to be either conformally or non-conformally coupled. The endpoint of the RG flow corresponds to a conformally coupled scalar field at the Wilson-Fisher fixed point. We take a spherical entangling surface in de Sitter space and compute the entanglement entropy everywhere along the RG trajectory. In $4$ dimensions, a free non-conformal  scalar has a universal area term scaling with the logarithm of the UV cutoff. In $4-\epsilon$ dimensions, such a term scales as $1/\epsilon$. For a non-conformal scalar, a $1/\epsilon$ term is present both at the UV fixed point, and its vicinity. For flow between two conformal fixed points, $1/\epsilon$ terms are absent everywhere. Finally, we make contact with replica trick calculations. The conical singularity gives rise to boundary terms residing on the entangling surface, which are usually discarded. Consistency with our results requires they be kept. We argue that, in fact, this conclusion also follows from the work of Metlitski, Fuertes, and Sachdev, which demonstrated that such boundary terms will be generated through  quantum corrections.

}

\begin{document}

\maketitle

\section{Introduction}
An important aspect of quantum field theory is the renormalization group (RG) flow between  conformal field theories \cite{Zamolodchikov:1986gt, Cardy:1988cwa, Komargodski:2011vj, Komargodski:2011xv}. Recent results, such as the proof of the F-theorem by Casini and Heurta \cite{Casini:2012ei, Casini:2004bw}, strongly suggest entanglement entropy plays an important role in characterizing such flows. For some recent studies of entanglement and RG, see \cite{Myers:2010xs, Giombi:2014xxa, Ben-Ami:2014gsa,Park:2015dia,Fei:2015oha, Perlmutter:2015vma, Casini:2015ffa, Goykhman:2015sga}
To date, most explicit field theoretic computations of entanglement entropy have focused on the vicinity of fixed points. It is the purpose of this paper to compute entanglement entropy along an entire RG trajectory.  We do this for the interacting vector $O(N)$ model at large $N$ in $4$ and $4-\epsilon$ dimensions.  

\begin{figure}
	\begin{center}
		\includegraphics[width=5cm,height=5.1cm]{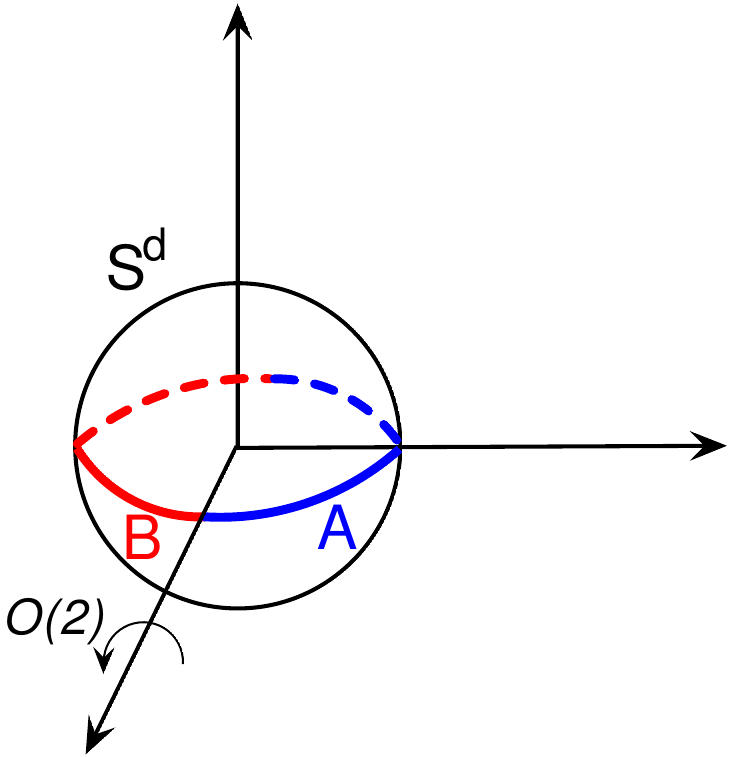}
		\caption{The equator of $S^d$ is split into two equal cap-like regions $A$ and $B$ (shown in blue and red respectively). The entanglement entropy is the von Neumann entropy of the density matrix living on the blue region.}
		\label{fig:sphere}
	\end{center}
\end{figure}

In Sec.~\ref{sec:review} we review the entanglement flow equations and recent results in perturbative calculations of entanglement entropy. We establish the setup of our problem: the Euclidean spacetime is taken to be a sphere, and the entangling surface is the bifrucation surface of the equator (see Fig. \ref{fig:sphere}). The radius $\ell$ of the sphere sets the RG scale. The flow equations for the variation of entanglement entropy with respect to $\ell$ reduce to a one-point function of the trace of the stress-tensor. 

In Sec.~\ref{model} we introduce the vector $O(N)$ model. Within the $\epsilon$ expansion, the RG flow is from the Gaussian fixed point in the UV to the Wilson-Fisher fixed point in the IR. We work to leading order in $1/N$, so that the dynamics is encoded in the mass gap equation. We take the theory to have arbitrary non-minimal coupling in the UV. We renormalize the theory,  solve for the beta functions, and  compute the expectation value of the trace of the stress-tensor.

In Sec.~\ref{sec:EE} we find the entanglement entropy as a function of $\ell$. This follows immediately from the results of the previous sections; the general expression is presented in Sec.~\ref{sec:RG}. 
It is instructive to first directly find the entanglement entropy in various limits, and in Sections \ref{sec:Gaussian}, \ref{sec:WF}, we study the UV and IR limits, respectively. In Sec.~\ref{sec:4D} we warm up with the case of strictly four dimensions; this does not have a UV fixed point, but one may still study the entanglement entropy between two points along the RG trajectory.

In Sec.~\ref{sec:bndry} we discuss the implications of our results for computations on backgrounds with conical deficits.  The entanglement flow equations imply entanglement entropy is sensitive to the amount of non-minimal coupling to gravity, even in the flat space limit and with interactions. We show this is consistent with replica-trick calculations, but only if one accounts for the contribution of the conical singularity. In fact, this conclusion is unavoidable: even if one chooses to discard such boundary terms, quantum corrections will generate them \cite{Sachdev}.

\section{Review of Entanglement Flow Equations} \label{sec:review}
In this Section we review the entanglement flow equations. The main equation, Eq.~\ref{main}, expresses the entanglement entropy in terms of the expectation value of the trace of the stress-tensor. 

Entanglement entropy is the von Neumann entropy of a reduced density matrix, $\rho$. The entanglement (or modular) Hamiltonian is defined through the reduced density matrix as 
\be
\rho = e^{-K}~, \quad \text{Tr}\,\rho=1~.
\ee
It trivially follows that entanglement entropy is given by the expectation value of the entanglement Hamiltonian, 
\be \label{eq:SK}
S_{\mt{EE}} \equiv - \text{tr} (\rho \log \rho) = \langle K \rangle~.
\ee
Viewing the expectation value in (\ref{eq:SK}) through the Euclidean path integral representation, the entanglement flow equations \cite{Rosenhaus:2014nha, Rosenhaus:2014woa}  follow:
\bea \label{flow} 
\frac{\partial S_{\mt{EE}}}{\partial \lambda} &=& - \int \langle  \mathcal{O}(x) \, K\rangle\ \\ \label{flowg}
2\frac{\delta S_{\mt{EE}}}{\delta g^{\mu \nu} (x)} &=& -\sqrt{g(x)} \langle  T_{\mu \nu}(x) \, K\rangle~,
\eea
where the integral runs over the entire Euclidean manifold parametrized by $x$.
These equations describe the change in entanglement entropy under a deformation of the coupling $\lambda$ of some operator $\mathcal{O}$, or under a change in the background metric $g_{\mu \nu}$.\footnote{Eq.~\ref{flowg} can also be used to study shape dependance of entanglement entropy \cite{Rosenhaus:2014woa}, although there are unresolved issues at second order \cite{Rosenhaus:2014zza}. Shape dependance will not be the focus of our study, see however \cite{solo,Allais:2014ata, Mezei:2014zla, Lewkowycz:2014jia, Carmi:2015dla, Faulkner:2015csl, Bianchi:2015liz}.}

A planar entangling surface in Minkowski space, or a spherical entangling surface in de Sitter space, are especially useful contexts in which to study the flow equations. In these two cases the symmetry in the transverse directions to the entangling surface  implies that the entanglement Hamiltonian is proportional to the boost generator (or rotation generator, in the Euclidean continuation) \cite{Kabat:1994vj} 
\be \label{eq:2}
K = \int_A T_{\mu \nu} \xi^{\mu} n^{\nu} + c~,
\ee
where  $n^{\nu}$ is normal to the entangling surface, $\xi^{\mu}$ is the Killing vector associated with the symmetry, and $c$ is a normalization constant such that $\text{Tr}\,\rho=1$. It should be noted that Eq.~\ref{eq:2} is valid for any Lorentz invariant quantum field theory. 

Let us review a few properties of the flow equation (\ref{flow}). Since the correlator $\langle T_{ab} \mathcal{O}\rangle$ vanishes for a CFT, the change in entanglement entropy under a deformation away from a fixed point vanishes to first order in $\lambda$ \cite{Rosenhaus:2014woa, Rosenhaus:2014ula}. This demonstrates the stationarity of entanglement entropy at the fixed points on a sphere, providing an affirmative answer to the question raised in \cite{Klebanov:2012va}.\footnote{In order to see stationarity of entanglement entropy at the conformal fixed points on a sphere, it should be the case that the stress-tensor in the correlator $\langle T_{ab} \mathcal{O}\rangle$ represents CFT degrees of freedom only. In later sections, we will find that when the IR fixed point is reached by an RG flow from the UV, this is not necessarily the case. In our case the stress-tensor for the gapped system does not vanish in the deep IR and gives rise to nonzero entanglement entropy which can be regarded as a remnant of the UV.} The distinction between the conformally and nonconformally coupled scalar is something we will return to. The second order in $\lambda$ part of entanglement entropy is fixed by the correlators $\langle T_{a b} \mathcal{O} \mathcal{O} \rangle$ and $\langle \mathcal{O} \mathcal{O}\rangle$, and is thereby completely universal: the result agrees with both free field and holographic computations \cite{Rosenhaus:2014zza}. And while these calculations are done for a planar entangling surface, the result for this universal
entanglement entropy ($\text{log}$ term)
is independent of the shape of the entangling surface \cite{Rosenhaus:2014zza}, as was verified holographically \cite{Taylor}. 

For a planar entangling surface, one can give an independent derivation of (\ref{flow})  \cite{Rosenhaus:2014ula}. In addition, through use of spectral functions, one can give a compact expression for the entanglement entropy for a general QFT \cite{Rosenhaus:2014ula}, allowing a demonstration of the equivalence of entanglement entropy and the renormalization of Newton's constant \cite{Casini:2015aa,Casini:2015ffa}. 

While a planar entangling surface is a useful and simple case to consider, it is a bit too simple for the study of RG flow of entanglement entropy, as it lacks any scale. A spherical entangling surface is the best suited in this regard, as the size of the sphere sets the RG scale. For an entangling surface that is a sphere in flat space,  the entanglement Hamiltonian is only known for a CFT \cite{Casini:2011kv}, which is sufficient for computing entanglement entropy perturbatively near the fixed point \cite{Faulkner:2014jva}, but insufficient for finding it along an entire RG trajectory. 

In this paper, we study entanglement entropy for a spherical entangling surface in de Sitter space. In this case, one knows the entanglement Hamiltonian along the entire RG trajectory, and the flow equation can be directly applied. The analytic continuation of de Sitter is a sphere $S^d$, and the Killing vector $\xi^{\mu}$ is the rotation generator. We will study the change of entanglement entropy as we vary the radius $l$ of $S^d$.  

Noting that variation of the radius $\ell$ of the space $S^d$ can be expressed as, 
\be
\ell \frac{\partial}{\partial \ell} = - 2 \int g^{\mu \nu}(x) \frac{\delta}{\delta g_{\mu \nu}(x)}~,
\ee
the flow equation (\ref{flowg})  gives,
\be \label{eq:flow1}
\ell\frac{d S_{\mt{EE}}}{d \ell}  = \int \langle T_{\mu}^{\mu}(x)\, K \rangle~.
\ee
As a result of  (\ref{eq:2}), the flow of entanglement entropy can  be computed from (\ref{eq:flow1}) provided one knows the 2-pt function of the stress-tensor. In fact,  a further simplification  can be made. As a result of the maximal symmetry of de Sitter space, as well as  the Ward identities, the 2-pt function can be reduced to a 1-pt function, turning (\ref{eq:flow1}) into \cite{Ben-Ami:2015aa}
\begin{equation}
\ell\frac{d S_{\mt{EE}}}{d \ell}  = - { V_{S^d} \over d} \,\ell{d\over d\ell} \langle T^\mu_\mu\rangle ~,
\label{main}
\end{equation}
where $V_{S^d}$ is the volume of a $d$-dimensional sphere of radius $\ell$. Eq.~\ref{main} can also be found directly from the interpretation of entanglement entropy as the thermal entropy in the static patch \cite{Ben-Ami:2015aa}. The flow equation in the form (\ref{main}) will be used in Sec.~\ref{sec:EE} to compute entanglement entropy throughout the RG flow, from $l\rightarrow 0$ in the UV to $l\rightarrow \infty$ in the IR.

\section{$O(N)$ on a sphere} 
\label{model}
In this Section we introduce the field theory background for the $O(N)$ model on a sphere. We work within the $\epsilon$ expansion, so as to have both the free UV and the Wilson-Fisher IR fixed points. We also work at large $N$, allowing us to sum the infinite class of cactus diagrams, as is concisely encoded in the mass gap equation. For the purposes of computing the $\beta$ functions near the fixed points, this is equivalent to working at finite $N$ to one-loop. 

After introducing the action and the gap equation in Sec.~\ref{modelIntro}, we renormalize the theory and compute the $\beta$ functions in Sec.~\ref{sec:beta}, and find the expectation value of the trace of the stress-tensor in Sec.~\ref{sec:stress}. 

\subsection{Gap Equation} \label{modelIntro}
The Euclidean action of the $O(N)$ vector model living on a $d$-dimensional sphere of radius $\ell$ is given by, \footnote{We do not distinguish between the bare and renormalized $\phi$ since to leading order in $1/N$ they are identical.}
\be \label{action}
I = \int_{S^d} \left[ \frac{1}{2} (\partial \vec\phi)^2 + {t_0 \over 2} \vec \phi^2
 +{1\over 2}(\xi\eta_c+\eta_0) R \vec \phi^{\,2} 
 +\frac{u_0}{4 N}(\vec \phi ^{\,2})^{2}\right] +I_g~,
 \ee
where $R={d(d-1)\over \ell^2}$ is the scalar curvature and $\eta_c={d-2\over 4(d-1)}$ is the conformal coupling.\footnote{The bare coupling $\eta_0$ is introduced to account for the possible counter-terms associated with renormalization of the non-minimal coupling to gravity.} We take the theory to have arbitrary non-minimal coupling, parameterized by $\xi$; we will be interested in letting $\xi$ have the range $ \xi \geq 0$, with $\xi = 0$ corresponding to a minimally coupled scalar and $\xi = 1$ corresponding to a conformally coupled scalar. The lower bound on $\xi$ follows from the requirement that the theory is stable in the UV. 

Since we are on a curved space, we have included $I_g$ which describes the purely gravitational counter-terms which must be introduced to cancel the  vacuum fluctuations of $\vec{\phi}$,
\be \label{eq:Ig}
I_g = N\int_{S^d} \left[ \Lambda_0 + \kappa_0 R + a_0 E_4 + c_0 R^2 \right],
\ee
where $R$ is the Riemann scalar and $E_4=R_{\al\bt\gamma\delta}R^{\al\bt\gamma\delta} - 4 R_{\al\bt}R^{\al\bt} + R^2$.\footnote{In 4-dimensions $E_4$ represents the Euler density. Also, since the sphere is conformally flat, there is no  Weyl tensor counterterm.} 

Following the standard large $N$ treatment, we introduce a Lagrange multiplier field $s$ and an auxiliary field $\rho$, so as to write the generating function as
 \be
 Z[\vec{J}] = \int \mathcal{D} s\ \mathcal{D} \rho\ \mathcal{D} \vec{\phi}\ \exp\( - I - \vec{J}\cdot \vec{\phi}\)~,
 \ee
 where the action is
\be 
 I=\int \left[\frac{1}{2} (\partial \vec \phi)^2 + {t_0 \over 2} \vec \phi^2
 +{1\over 2} \, \xi \eta_c R\ \vec \phi^{\,2} +{1\over 2} s(\vec \phi^{2}-N \rho)
 +{N\over 2} \eta_0 R \rho + N \frac{u_0}{4}\rho^{2}\right] + I_g~.
 \ee
We can integrate out $\vec{ \phi}$ to obtain, \footnote{There is no spontaneous symmetry breaking on a sphere, so we are always in the $O(N)$ symmetric phase.}
\be
 Z[\vec{J}]  =\int \mathcal{D} s\ \mathcal{D} \rho\
 \exp\Big( -\frac{N}{2} \int \(\frac{u_0}{2}\rho ^2-s \rho + \eta_0 R \rho\) -\frac{N}{2}  \text{Tr} \ln \hat{O}_{s} +\frac{1}{2} \langle\vec{J}, \hat O^{-1}_{s} \vec{J}\rangle - I_g\Big)
 \, ,
 \ee
 where $\hat{O}_{s}\equiv -\square + t_0 + \xi\eta_cR +s$ and $\langle,\rangle$ is the $L^2$ norm.  The contours of integration for the auxiliary fields $s$ and $\rho$ are chosen so as to ensure that the path integral converges. 
 
The auxiliary field $\rho$ has trivial dynamics since it appears algebraically in the action. It can  easily be integrated out, 
 \be
 Z[{\vec{J}}]  =\int \mathcal{D} s \,
 \exp\Big( -\frac{N}{2} \( \text{Tr} \ln \hat{O}_{s} - \int_{S^d} {(s-\eta_0 R)^2\over 2 u_0} \)  +\frac{1}{2} \langle\vec{J}, \hat O^{-1}_{s} \vec{J}\rangle  - I_g  \Big)
 \, ,
 \labell{part}
 \ee
The remaining auxiliary field $s$ encodes the full dynamics of the original $N$ physical degrees of freedom; it is an $O(N)$ singlet, which significantly simplifies the $1/N$ expansion.  
At large $N$ the theory is dominated by the saddle point, $s(x) = \bar s$, which satisfies
 \be \label{saddle}
  \bar s= u_0 \langle\phi^2\rangle + \eta_0 R~, \quad
  \langle\phi^2\rangle\equiv \langle x| \hat{O}_{\bar s}^{\,-1} |x\rangle ~.
 \ee
It is convenient to re-express (\ref{saddle}) in terms of the physical mass,\footnote{If the space-time is flat, then $m^2$ corresponds to a pole in the 2-point function.} $m^2\equiv t_0+\bar s$,
\be
 m^2 =  t_0 + u_0 \, \langle\phi^2\rangle + \eta_0 R ~.
 \labell{gapeq}
\ee

\begin{figure}[tbp] 
\centering
\subfigure[]{
	\includegraphics[width=4in]{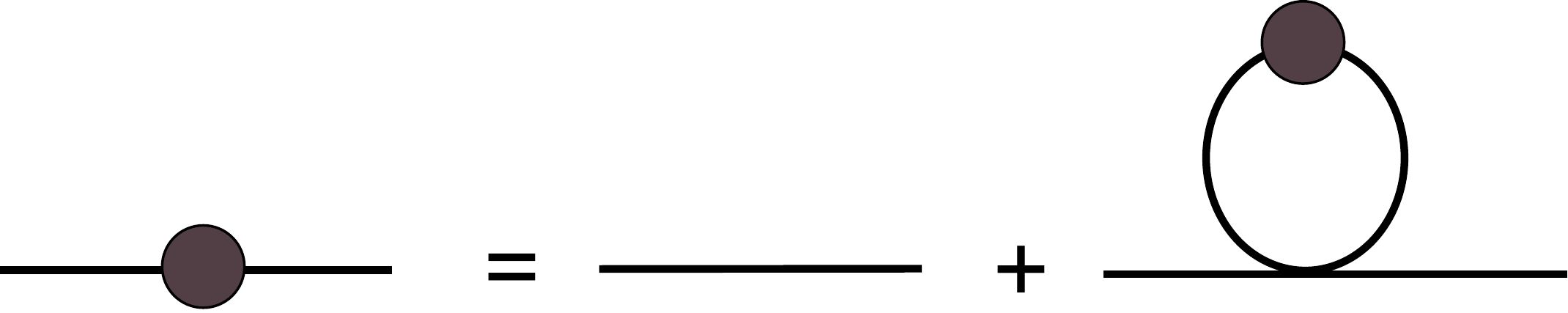}
	}
		
		\subfigure[]{
	\includegraphics[height=1.3in]{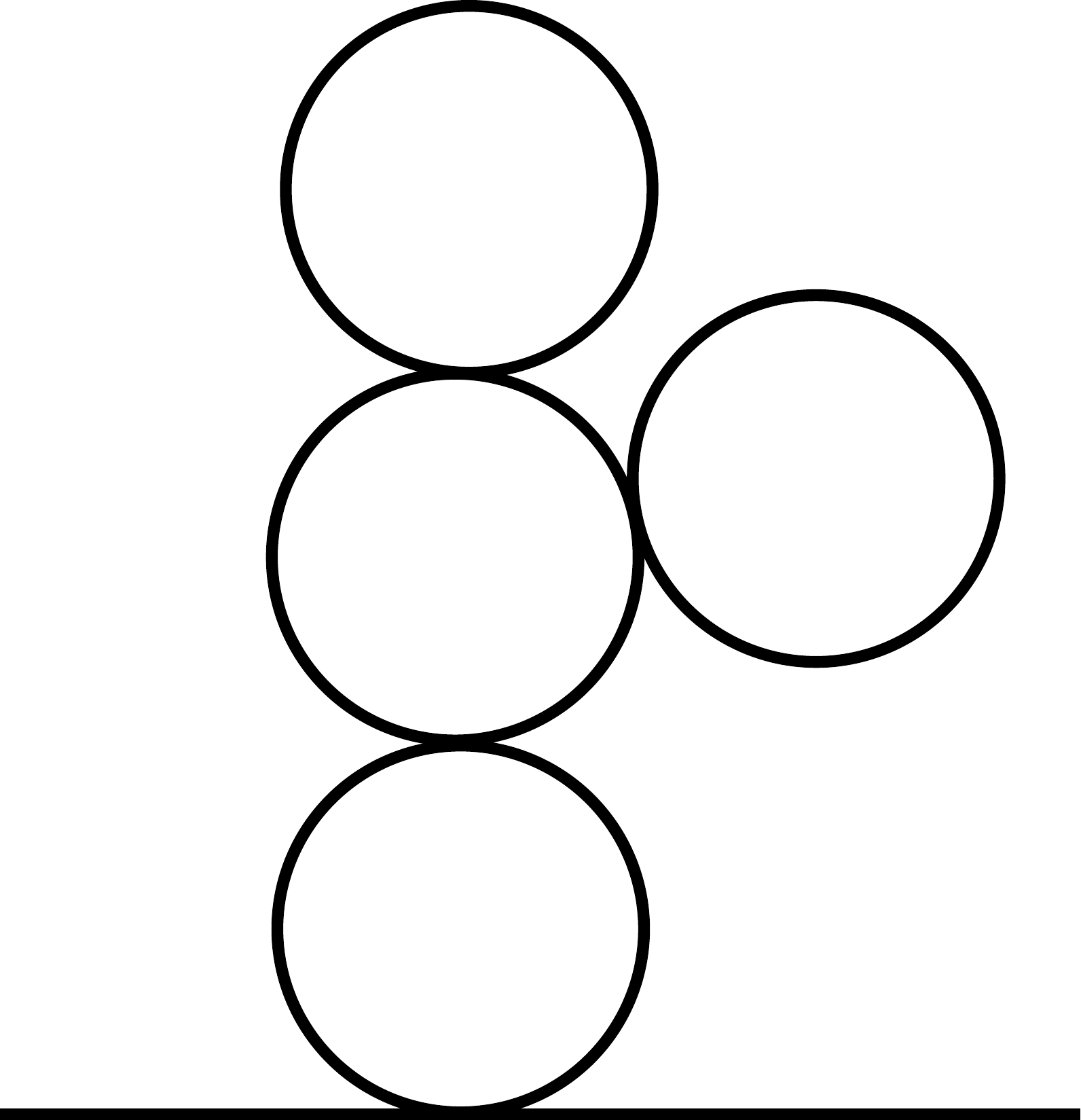}
	}
\caption{(a) At large $N$ the Schwinger-Dyson equation for the 2-pt function simplifies to become the gap equation (\ref{gapeq}). (b) Iterating (a) gives a sum of cactus diagrams, such as the one shown above.}
\label{fig:gap}
\end{figure}

Eq.~\reef{gapeq} is the gap equation; it has a simple interpretation. At large $N$, fluctuations of $\vec{\phi}^2$ are suppressed, $\langle \vec{\phi}^2(x) \vec{\phi}^2(y)\rangle \approx \langle \vec{\phi}^2(x)\rangle \langle\vec{\phi}^2(y)\rangle$. The quartic interaction in the action (\ref{action}) is thus effectively the square of a quadratic, and at leading order in $1/N$ the theory can in some sense be regarded as a free theory with the mass fixed self-consistently through (\ref{gapeq}). Equivalently, at large $N$ the propagator is found by summing over all cactus diagrams (see Fig.~\ref{fig:gap}); this sum is encoded in (\ref{gapeq}), as can be seen by iterating (\ref{gapeq}) starting with the bare mass $t_0$. 

To solve the gap equation (\ref{gapeq}) we need the two-point function  on a sphere for  a free field of mass squared $m^2 + \xi \eta_c R$,
\be \label{2pt}
\langle \phi_a(x) \phi_c(y)\rangle =\delta_{ac} \, \frac{l^{2-d}}{(4\pi)^{d/2} }\frac{\Gamma(\lambda) \Gamma(-\lambda + d-1)}{\Gamma(d/2)} \ _2 F_1(\lambda, d-1-\lambda, \frac{d}{2}; \cos^2\frac{\chi}{2}) ~,
\ee
where $\chi$ is the angle of separation between $x,y$ and 
\be \label{eq:lambda}
\lambda={d-1\over 2 } + i \sqrt{(m\ell)^2-{1\over 4} +  {d(d-2)(\xi-1)\over 4} }~.
\ee
In the limit of coincident points (\ref{2pt}) becomes
\be 
\langle\phi^2\rangle=
 {\Gamma(1-d/2)\Gamma(\lambda)\Gamma(d-1-\lambda)\over \pi (4\pi)^{d/2}\ell^{d-2}} \sin\big(  {\pi\over 2}(d-2\lambda)  \big)\, , 
  \labell{phi2}
\ee
Eq.~\reef{phi2} exhibits a logarithmic divergence in the vicinity of $d=4$, and so we must renormalize the theory. 

\subsection{Beta functions to leading order in $1/N$} \label{sec:beta}
We first consider the counter terms needed to renormalize the couplings $t, u, \eta$ in the action for $\phi$. This is done through use of the gap equation and the requirement of a finite mass $m$. 

The divergent piece in the gap equation (\ref{gapeq}) can be obtained by expanding \reef{phi2} in $\epsilon\equiv4-d\ll 1$,
\be
\langle\phi^2\rangle = \({(1-\xi) \over 4\pi^2\ell^2} - {m^2 \over 8\pi^2}  \){m^{-\epsilon} \over \epsilon} +\mathcal{O}(\epsilon^0)~.
\labell{sing2p}
\ee
To ensure a finite mass gap, $m^2$, the bare parameters $u_0,  t_0$ and $\eta_0$ in Eq.~\reef{gapeq} should be renormalized. To find the relation between the bare and renormalized couplings we rewrite Eq.~\reef{gapeq} as,
\be
 {m^2\over u_0} = { t_0\over u_0} + {\eta_0\over u_0} R - {m^2 \ell^{\epsilon}\over 8\pi^2\epsilon} + {(1-\xi)\ell^{\epsilon} \over 4\pi^2\ell^2 \epsilon}  +\ldots ~,
 \labell{massgap2}
\ee
where ellipsis encode finite terms independent of the bare couplings. The absence of poles in the gap equation therefore gives the following relation between the bare and renormalized parameters:
\be
 {1\over u_0}={1\over u  \mu^\epsilon}-{1\over 8\pi^2\epsilon\, \mu^\epsilon} ~, \quad\quad { t_0\over u_0}={t \over u \,\mu^\epsilon}~,
 \quad\quad {\eta_0\over u_0} ={\eta \over u \mu^\epsilon} - {1-\xi\over 48\pi^2 \epsilon\,\mu^\epsilon}~,
 \labell{renorm}
\ee
where $t, u, \eta$ are the renormalized variables and depend on the RG scale $\mu$.\footnote{The couplings $u $ and $\eta $ are dimensionless, and we use the minimal subtraction scheme throughout the paper.}

Applying $\mu {d\over d\mu}$ to both sides of \reef{renorm}, and recalling that the bare couplings are independent of $\mu$, leads to the following set of RG equations:
\bea
 \beta_u&\equiv&\mu {d u \over d\mu} =-\epsilon u  + {u ^2\over 8\pi^2}={u \over 8\pi^2}(u -u^*)~, 
 \non
 \beta_t&\equiv& \mu {d t  \over d \mu} = {u  t \over 8\pi^2}~,
  \label{beta}
  \\
 \beta_\eta&\equiv&\mu {d \eta \over d\mu}={u  \over 8\pi^2} \(\eta  - {1-\xi\over 6}\) ~.
\nonumber
 \eea
Here $u^*=8\pi^2\epsilon$  is the well-known Wilson-Fisher IR fixed point\footnote{At finite $N$ the Wilson-Fisher fixed point is at $u^* =N \frac{8 \pi^2 \epsilon}{N+8} = 8\pi^2\epsilon + O(1/N)$. Note that as a result of the normalization of the quartic term in the lagrangian by a factor of $N$, $u^*$ has an additional factor of $N$ as compared to the usual $\epsilon$ expansion conventions.}, while the Gaussian UV fixed point is at $u =0$. 

Our discussion has been general, in that we have allowed the theory to have an arbitrary non-minimal coupling in the UV. Given our conventions for the coefficient of non-minimal coupling to gravity, $(\xi \eta_c + \eta_0) R\vec{\phi}^2$ in the action (\ref{action}), we impose that $\eta=0$ at the UV fixed point. Since the constant $\eta_c$ was picked to be $\eta_c =1/6$ in $4$ dimesnions, this ensures that for $\xi =1$ the scalar is conformally coupled in the UV, while for $\xi=0$ it is minimally coupled. In fact, \reef{beta} tells us that independent of our choice of $\xi$, the IR endpoint of the flow is the same:  $\eta ^*= {1-\xi\over 6}$ at the Wilson-Fisher fixed point, and therefore $\xi\eta_c + \eta ^*=1/6$.
Thus to leading order in $\epsilon$, a family of weakly interacting, non-conformally coupled massive scalar fields, parametrized by $\xi\geq 0$ in the vicinity of the Gaussian fixed point, all flow to the conformally coupled theory at the Wilson-Fisher fixed point.

\subsection{Gravitational counter-terms and energy-momentum tensor}
The $\beta$ functions for $u$ and $t$ found in (\ref{beta}) are obviously the same as those in flat space. In addition to $\eta$, the sphere background requires the introduction of  gravitational counterterms (\ref{eq:Ig}). In this section we compute their $\beta$ functions; the expectation value of the trace of the energy-momentum tensor will then  follow. We note that knowing  the contribution of the gravitational counterterms is essentially irrelevant for our purposes. We are interested in the area law piece of entanglement entropy, which by necessity involves the mass. The contribution of the gravitational counterterms $a E_4$ and $c R^2$ only involves the sphere radius $\ell$ and correspondingly will give some constant contribution to the entanglement entropy. The computation of these conterterms is  for completeness; the reader uninterested in the details may skip to the result, Eq.~\reef{trace2}.

Our analysis and notation will closely follow the discussion in Ref.~\cite{Brown:1980qq}. We work on an arbitrary conformally flat curved background,\footnote{Conformally flat because we do not bother to include the Weyl tensor counter term.} specializing to a sphere at the end. 
The relation between the bare and renormalized parameters is \cite{Brown:1980qq}
\bea
\eta_0&=& (\eta+ F_\eta)Z_2^{-1} ~,\quad Z_2^{-1}\equiv {t_0\over t} ~,
\non
\Lambda_0 &=& \mu^{d-4}(\Lambda + t_0^2 F_{\Lambda}) ~,
\non
\kappa_0 &=& \mu^{d-4}(\kappa + 2 t_0 \eta_0 F_{\Lambda} + t_0 F_{\kappa})~,
\non
a_0 &=& \mu^{d-4}(a + F_a)~,
\non
c_0 &=& \mu^{d-4}(c + \eta_0^2 F_{\Lambda} + \eta_0 F_{\kappa} + F_c) ~,
\label{renorm2}
\eea
where all counter terms, $Z_2^{-1}, F_\eta, F_\Lambda, F_\kappa, F_a$ and $F_c$, are dimensionless functions of the renormalized coupling $u$, and we choose a scheme where they contain only an ascending series of poles in $\epsilon=4-d$. As argued in \cite{Brown:1980qq}, these functions are independent of the renormalized coupling $\eta$. Furthermore, from \reef{renorm} we immediately find,
\bea
 Z_2^{-1}&=& \(1-{u \over 8\pi^2\epsilon}\)^{-1} ~,
 \non
 F_\eta&=&{(\xi-1)u\over 48\pi^2\epsilon} ~.
 \label{Feta}
\eea
To calculate $F_\Lambda$ and $F_\kappa$, one can use the definition of the renormalized operator $[\vec\phi^{\,2}]$ \cite{Brown:1980qq}: \footnote{In our case the relative sign of the counter terms is flipped since we are using Euclidean signature.}
\bea
t\lbrack \vec\phi^{\,2}\rbrack=t_0\vec\phi^{\,2} + N\mu^{-\epsilon} t_0 \Big(4 (t_0+\eta_0R) F_{\Lambda}+2RF_{\kappa}\Big) ~,
\eea
and require that its vev is finite. The result is,
\bea
F_\Lambda&=&{1\over 2(4\pi)^2 \epsilon} \(1-{u \over 8\pi^2\epsilon}\) ~,
\non
F_\kappa&=&{\xi-1\over 96\pi^2\epsilon}\( 1- {u\over 8\pi^2\epsilon}\) ~.
\label{FLamFk}
\eea
In Appendix \ref{Flam} we carry out an independent calculation of $F_\Lambda$, finding agreement with \cite{Fei:2015oha} and the above result. Using these counter terms together with \reef{renorm2}, one can  evaluate the RG flow equations for $\Lambda$ and $\kappa$,
\bea
 \mu {d\Lambda \over d\mu} &=& \epsilon\Lambda + {t^2\over 2(4\pi)^2} ~,
 \non
 \mu {d\kappa \over d\mu} &=&  \epsilon\kappa + {t  \over (4\pi)^2} \(\eta - {1-\xi\over 6}\) \, ~.
 \labell{RGkappa}
\eea
The RG equation for $\kappa$ can then be solved,
\be
 \kappa={\eta\over 2u} \, t + \bar{\kappa} \mu^{\epsilon},
 \labell{kappa}
\ee
where $\eta^*=(1-\xi )/6$ and $\bar{\kappa}$ is some constant that will still need to be fixed.

Turning to the $a$ and $c$ coefficients, 
\bea
 \beta_a&\equiv&\mu{d a\over d\mu} = \epsilon \, a + f_a^{(1)}~,
 \non
 \beta_c&\equiv&\mu{dc\over d\mu} =\epsilon \, c + {\eta^2\over 2(4\pi)^2} + \eta {\xi-1\over 96\pi^2} + f_c^{(1)}~,
\eea
where $f_a^{(1)}$ and $f_c^{(1)}$ are the residues of the simple poles in the definitions of $F_a$ and $F_c$, 
\bea
 F_a&=& {f_a^{(1)} \over \epsilon} + {f_a^{(2)} \over \epsilon^2} + \ldots ~,
 \non
 F_c&=& {f_c^{(1)} \over \epsilon} + {f_c^{(2)} \over \epsilon^2} + \ldots~.
\eea
To calculate $F_a$ and $F_c$ we will require that the energy-momentum tensor has a finite vev. 

\subsubsection*{Energy-momentum tensor} \label{sec:stress}

The energy-momentum tensor of the $O(N)$ model is given by $T_{\mu\nu}=T_{\mu\nu}^\phi+T_{\mu\nu}^g$ where $T_{\mu \nu}^\phi$ is the contribution from $\phi$,
\bea 
T_{\mu\nu}^\phi&\equiv&{2\over\sqrt{g}}{\delta I_g\over \delta g^{\mu\nu}}=
\nabla_\mu\vec\phi\nabla_\nu\vec\phi - g_{\mu\nu} \( {1\over 2}(\del\vec\phi)^2 + {1\over 2}  t_0 \vec\phi^{\,2} 
+ {1\over 2} \xi\eta_c R \, \vec\phi^{\,2}  +\frac{u_0}{4 N}(\vec\phi ^{\,2})^{2}\) 
\non 
&&\quad\quad\quad\quad+\xi\eta_c R_{\mu\nu}\vec\phi^{\,2} + \xi\eta_c\(g_{\mu\nu}\nabla^2-\nabla_\mu\nabla_\nu\)\vec\phi^{\,2} ~, \nonumber
\non
\eea
and $T_{\mu \nu}^g$ is the gravitational contribution, defined through the variation of the action, 
\be
T_{\mu\nu} ={2\over\sqrt{g}}{\delta I_g\over \delta g^{\mu\nu}}.
\ee
Taking the trace gives
\be
T\equiv g^{\mu\nu}T_{\mu\nu}=- t_0 \vec\phi^{\,2} -{d-2\over 2} E_0 + (d-4) {u_0\over 4N} (\vec\phi^{\,2})^2+(d-1)(\xi-1)\eta_c\nabla^2\vec\phi^{\,2}+T^g~,
\ee
where $E_0$ is the equation of motion operator 
\be
 E_0=\vec\phi\(-\nabla^2+\xi\eta_c R+ t_0+\frac{u_0}{N}(\vec\phi ^{\,2})\)\vec\phi ~,
\ee
and $T^g$ is the trace of the gravitational part, 
\be
 {1\over N}T^g \equiv {1\over N}g^{\mu\nu}T_{\mu\nu}^g=-d\Lambda_0-(d-2)\kappa_0 R - (d-4)(a_0E_4+c_0R^2) ~,
\ee
where we dropped a term proportional to $\nabla^2R$, since it vanishes on a sphere.

Taking the vev of the energy-momentum trace and using the gap equation (\ref{gapeq}): $\langle \vec\phi^2 \rangle = N (m^2-t_0-\eta_0 R)/u_0$,  yields\footnote{Note that the vev of the equation of motion operator vanishes identically. The same holds for vevs of total derivatives on a sphere.}
\be
 {1\over N}\langle T \rangle=- \, {t_0\over u_0} \, (m^2-t_0-\eta_0 R)  + (d-4)  \, {1\over 4 u_0} \, (m^2-t_0 - \eta_0 R)^2+{1\over N}T^g~.
 \labell{trT}
\ee
By definition, this expression is finite after the bare parameters are expressed in terms of the renormalized parameters. Substituting \reef{renorm}, \reef{renorm2}, \reef{Feta} and \reef{FLamFk} into \reef{trT}, we get after some algebra,
\bea
 {\mu^\epsilon\over N}\langle T \rangle&=&{ m^4 \over 2 (4\pi)^2} \(1- {u^*\over u }\) + { t  \over u } \Big(t -m^2+{\epsilon\over 2} \Big(m^2-{t \over 2}\Big)\Big) 
 - (4-\epsilon)\Lambda 
 \non
  &+&  \((\epsilon -2)\kappa+\Big(1-{\epsilon\over 2}\Big) {\eta\over u} \, t+\Big(\epsilon \,{\eta\over 2 u} + {\xi-1\over 96\pi^2}\Big)m^2\) R 
   \label{trace}
\\
  &+&\epsilon(a+F_a) E_4  + \epsilon \(c-{\eta^2\over 4u}+F_c+\Big({\xi-1\over 96\pi^2\epsilon}\Big)^2u\)R^2
  +\mathcal{O}(\epsilon^2)~.
  \nonumber
\eea
Imposing that $\langle T \rangle$ be finite leads to the following large-$N$ results:
\bea
 F_a&=& {f_a^{(1)} \over \epsilon}  ~,
 \non
 F_c&=& {f_c^{(1)} \over \epsilon} -\Big({\xi-1\over 96\pi^2}\Big)^2{u\over \epsilon^2}~.
\eea
Both $f_a^{(1)}$ and $f_c^{(1)}$ are not fixed. However,  to leading order in $1/N$ they are identical to their free field values, and we get
\bea
 F_a&=& \frac{-1}{360(4\pi)^2 \epsilon} ~,
 \non
 F_c&=& -\Big({\xi-1\over 96\pi^2}\Big)^2{u\over \epsilon^2}~.
\eea

Combining with \reef{kappa}, the final expression for the trace of energy-momentum tensor takes the form
\bea
 {\mu^\epsilon\over N}\langle T \rangle&=&{ m^4 \over 2 (4\pi)^2} \(1- {u^*\over u }\) + { t  \over u } \Big(t -m^2+{\epsilon\over 2} \Big(m^2-{t \over 2}\Big)\Big) 
 - (4-\epsilon)\Lambda 
 \non
  &+&  \((\epsilon -2)\bar\kappa\mu^\epsilon+\Big(\epsilon \,{\eta\over 2 u} + {\xi-1\over 96\pi^2}\Big)m^2\) R 
\label{trace2}
\\
  &+&\epsilon\(a-\frac{1}{360(4\pi)^2 \epsilon}\) E_4  + \epsilon \(c-{\eta^2\over 4u}\)R^2
  +\mathcal{O}(\epsilon^2)~,
  \nonumber
\eea
where recall that $R$ is the curvature of the sphere, the renormalized couplings are evaluated at RG scale $\mu$, the constant $\xi$ parameterizes the non-minimal coupling ($\xi=1$ for conformally coupled), and $m$ is the physical mass found through the gap equation. In  the next section, we use \reef{trace2} to calculate entanglement entropy.

\section{Entanglement entropy}
\label{sec:EE}

Having assembled the necessary field theory ingredients in the previous section, we now compute the entanglement entropy. In what follows we account for the leading order in $1/N$ contributions. 

The entanglement entropy is found by solving the flow equation (\ref{main}),  which involves the derivative of $\langle T\rangle$ \reef{trace2} with respect to the sphere radius $\ell$. Since the renormalized couplings $t, u, \eta, \kappa$ are independent of $\ell$, \footnote{The couplings of local interactions should not know about the global geometry.} we get 
\be
 \ell {dS_\mt{EE}\over d\ell}=-{2\pi\,N\,A_\Sigma\,\mu^{-\epsilon}\over (4-\epsilon)(3-\epsilon)}  \ell^2
 \( { m^2 \over  (4\pi)^2} \(1- {u^*\over u }\)\ell {dm^2\over d\ell}  +{t\over u}\({\epsilon\over 2} - 1\) \ell {dm^2\over d\ell} - 2(\epsilon -2) \bar\kappa \mu^{\epsilon} R \) + s_0 ~,
 \labell{main2}
\ee
where $A_\Sigma$ is the area of the entangling surface $\Sigma=S^{d-2}$, $s_0$ collectively denotes the contribution of curvature square terms in \reef{trace2}, and we used solutions of RG equations for $\kappa$ and $\eta/u$ to simplify the linear curvature term in \reef{trace2}, see \reef{kappa} and \reef{etau}. We ignore the $s_0$ terms as they are $m$-independent, and therefore just give some constant contribution to the entanglement entropy. The solution for  entanglement entropy along the RG trajectory follows upon substituting the solutions of the gap equation (\ref{gapeq}) and the beta functions (\ref{beta}) into \reef{main2} and integrating. 

There are, however, a few caveats associated with the standard ambiguities of renormalization.
Indeed, the couplings in the above expression depend on an arbitrary RG scale, $\mu$, as well as on the choice of renormalization scheme. This ambiguity is unsurprising, and reflects the well-known fact that entanglement entropy in field theory depends on the details of the regularization procedure, and is therefore  scheme dependent. However, certain contributions to the entanglement entropy, `universal entanglement entropy', are unaffected by a change in the regularization scheme. It is these terms that we will be interested in calculating. 

There are three competing scales: $\mu$, $\ell$, and the asymptotic mass $m_\infty$ given through the solution of the gap equation \reef{gapeq} in the limit of flat space.\footnote{It is apparent that $m$ should depend on $\ell$: the mass was found by summing cactus diagrams, which probe the entire sphere.}. Since the curvature of the sphere sets the characteristic energy scale, we must have $\mu\sim \ell^{-1}$. The constant of proportionality between $\mu$ and $\ell^{-1}$ is arbitrary, though this is no different than the usual freedom to rescale $\mu$. In our context, this constant of proportionality is exchanged for the entanglement entropy at some radius $\ell$. Or, put differently, we can express the entanglement entropy at radius $\ell_1$ in terms of the entanglement entropy at some other radius $\ell_0$. 

Furthermore, there is a substantial difference between the two cases characterized by $\epsilon=0$  and $\epsilon\neq 0$. In the former case the theory is not UV complete; as $m\ell$ runs from small to large values,  it flows from a nonconformal interacting field theory in the UV to a Gaussian IR fixed point. In contrast, for finite $\epsilon\ll 1$, the system flows  from the Gaussian UV fixed point at $m_\infty\ell\ll\epsilon\ll1$, into the interacting Wilson-Fisher IR fixed point at $m_\infty\ell\gg \epsilon^{-1}\gg 1$. In particular, there is no smooth limit which interpolates between the two cases, and we analyze them separately. We consider $\epsilon=0$ in Sec.~\ref{sec:4D}, and finite $\epsilon$ in Sec.~\ref{sec:WF},  \ref{sec:Gaussian}, and \ref{sec:RG}. Note that in 4-dimensions, the coefficient of the $\log$ term is universal. In $4-\epsilon$ dimensions there is no $\log$ term, however a $1/\epsilon$ term turns into a $\log$ in the limit of $\epsilon \rightarrow 0$.

\subsection{Four dimensions}
\label{sec:4D}
In this section we compute the entanglement entropy in $4$ dimensions. 
Taking the limit $\epsilon\to 0$ results in the following RG flow equations,
\bea
 \mu {d u \over d\mu} &=& {u ^2\over 8\pi^2}~, 
 \non
 \mu {d \over d\mu} \({\eta\over u}\)&=&  {\xi-1\over 48\pi^2 } ~.
\eea
Integrating gives,
\bea 
 u(\mu)&=&{u_0\over 1-{u_0\over 8\pi^2}\log(\mu\delta)} ~,
 \non
 {\eta\over u}&=& {\xi-1\over 48\pi^2 } \log\( \mu\delta \) ~,
 \labell{4Dsol}
\eea
where we imposed the initial conditions $(u,\eta)=(u_0, 0)$ at the UV scale $\delta$. In the deep IR  \reef{4Dsol} gives $(0, (1-\xi)/6)$. 

The 4D counterparts of \reef{trace2} and \reef{main2} are 
\bea
 {\langle T \rangle\over N}&=&{ m^4 \over 2 (4\pi)^2}  + { t  \over u } \Big(t -m^2\Big) 
 - 4\Lambda 
 -  \(2\bar\kappa+{1-\xi\over 96\pi^2}\,m^2\) R 
-\frac{E_4}{360(4\pi)^2}   ~,
\eea
and
\bea
   \ell {dS_\mt{EE}\over d\ell}&=&{\pi\,N\,A_\Sigma\over 6}  \ell^2
 \( {t\over u} \, \ell {dm^2\over d\ell} - 4 \bar\kappa R \) + s_0 ~,
 \labell{4DdS}
\eea
while the gap equation \reef{gapeq} can be succinctly written as 
\be
 {t\over u} = {m^2\over u(m)} - {\xi-1\over 4\pi^2\ell^2}\log(m\delta) - \langle \phi^2 \rangle_\text{reg}~,
 \labell{gapeq2}
\ee
where we used \reef{4Dsol}, and the last term denotes the regular part of the 4D two-point function in the limit of coincident points, see \reef{sing2p},
\be
 \langle \phi^2 \rangle_\text{reg} \equiv \lim_{\epsilon\to 0}\[\langle x| \( -\square + m^2 + \xi\eta_cR \)^{\,-1} |x\rangle -
  \({(1-\xi) \over 4\pi^2\ell^2} - {m^2 \over 8\pi^2}  \){m^{-\epsilon} \over \epsilon}\]~.
\ee
Applying a derivative with respect to $\ell$ to both sides of \reef{gapeq2}, yields
\be
  \ell {dm^2\over d\ell}\( 1+{u(m)\over 8\pi^2}\Big({1-\xi\over (m\ell)^2} - {1\over 2} \Big)  \)
 = u(m) \, {1-\xi\over 2\pi^2\ell^2} \log(m\delta) + u(m) \, \ell {d\over d\ell} \langle \phi^2 \rangle_\text{reg}
 \labell{dm2dL}
\ee
This expression together with \reef{4DdS} and \reef{gapeq2} provides a full solution for entanglement entropy flow on a sphere. Note that the RG scale $\mu$ is completely eliminated from the final answer. Effectively its role is played by the radius of the sphere, $\ell$, as the curvature of the background sets the characteristic energy scale for excitations. In particular, the deep IR and UV are defined by $m\ell\gg1$ and $m\ell\ll1$, respectively. Of course, we have assumed that the physical scales $m$ and $\ell$ are far away from the microscopic UV cut off $\delta$,
\be
 m\delta\ll 1  \quad \text{and}\ \quad  \ell\gg\delta ~.
 \labell{scales}
\ee

Now we explicitly evaluate EE in the UV and IR regimes. We start from the former. Using \reef{phi2} to evaluate 
$\langle\phi^2\rangle_\text{reg}$ and substituting the result into \reef{dm2dL}, yields\footnote{Note that based on \reef{scales} 
$u(m)\simeq -{8\pi^2\over \log(m\delta)}\ll 1$}
\be
\ell {dm^2\over d\ell}\Big|_{m\ell\ll 1}
 =u(m) {\xi-1\over 2\pi^2  \ell^2} \big(\log(m\ell)-\log(m\delta) + \ldots\big) +\mathcal{O}(u(m)^2)~,
\ee
where ellipsis denote subleading terms in $m\ell\ll1$. Similarly, from the gap equation \reef{gapeq2}, we get
\be
 u(m) {t(\mu)\over u(\mu)} \Big|_{m\ell\ll 1}= m^2 + 2 {\xi-1\over \ell^2} \, + \ldots ~.
\ee
Only the first term on the right hand side contributes to the `area law', while other terms are either subleading corrections or contribute to a constant term in the entanglement entropy. Hence,
\be \label{eq:mls}
   \ell {dS_\mt{EE}\over d\ell}\Big|_{m\ell\ll 1}= 
  N(\xi-1) \, {A_\Sigma\, m^2 \over 12\pi  } \log{\ell\over\delta}    - 8\pi\,N\,A_\Sigma \bar\kappa + \ldots ~,
 \ee
Similarly, in the IR regime we have
\be
   \ell {dS_\mt{EE}\over d\ell}\Big|_{m\ell\gg 1}= 
  N(1-\xi) \, {A_\Sigma\, m^2 \over 12\pi  } \log(m\delta)   - 8\pi\,N\,A_\Sigma \bar\kappa + \ldots ~.
  \labell{IR4DEE}
 \ee

The above behavior of entanglement entropy has a  simple physical interpretation. In the UV regime we have $1\ll {\ell\over\delta}\ll{1\over m\delta}$, and the universal `area law' of entanglement scales as $\ell^2\log(\ell/\delta)$, \ie there is a logarithmic enhancement relative to the standard growth $\sim\ell^2$. This enhancement persists as we increase $\ell$ until $\ell m\sim 1$ is reached. Effectively, the universal `area law' at $m\ell\sim 1$ is built from all the massive degrees of freedom which have almost decoupled at this point. As we continue increasing $\ell$, the universal `area law' continues growing like $\ell^2$ until the hierarchy of scales is reversed, $1\ll{1\over m\delta}\ll {\ell\over\delta}$, and the flow terminates at the IR fixed point. 

In particular, the logarithmic `area law' in the deep IR represents entanglement of UV degrees of freedom. It has nothing to do with the IR field theory, which is empty in our case. As was noted in \cite{Rosenhaus:2014woa, Rosenhaus:2014ula} (see Sec.~\ref{sec:review}), the universal entanglement entropy vanishes for a conformally coupled scalar. Setting $\xi=1$ in (\ref{IR4DEE}) and (\ref{eq:mls}) recovers this result.

\subsection{Wilson-Fisher fixed point} \label{sec:WF}

In this section we calculate the entanglement entropy at the Wilson-Fisher fixed point in $4-\epsilon$ dimensions. 
This requires evaluating the right hand side of \reef{main2}, which involves the derivative of $m^2$ with respect to $\ell$. 

We start by expanding \reef{phi2} in $m\ell\gg \epsilon^{-1}\gg 1 $, \footnote{We do not expand in $\epsilon$. Note also that $\mathcal{O}\big((m\ell)^{-4}\big)$ and higher order terms in \reef{propexp} are proportional to $\epsilon$, and therefore they do not contribute to the divergence of $\langle\phi^2\rangle$ when $\epsilon\to 0$.}

\be
\langle \phi^2 \rangle = { (4\pi)^{\epsilon\over2}\over (2\pi)^2\epsilon(\epsilon-2)}\Gamma\(1+{\epsilon\over2}\) m^{2-\epsilon} 
\(1-{(\epsilon-4)(\epsilon-2)(6(1-\xi)+\epsilon(3\xi-2))\over 24 \,(m\ell)^2} + \mathcal{O}\big((m\ell)^{-4}\big)  \)
 \labell{propexp}
\ee
Now using \reef{gapeq}, \reef{renorm} and \reef{propexp} results in
\be
 t^*=  { 2(4\pi)^{\epsilon\over2}\over (2-\epsilon)}\Gamma\(1+{\epsilon\over2}\) m^2 \( {m\over\mu} \)^{-\epsilon}
\(1-{(\epsilon-4)(\epsilon-2)(6(1-\xi)+\epsilon(3\xi-2))\over 24 \,(m\ell)^2} + \mathcal{O}\big((m\ell)^{-4}\big)  \) ~,
 \labell{tstar}
\ee
where the asterisk in $t^*$ denotes that the system sits at the Wilson-Fisher fixed point. Differentiating \reef{tstar} with respect to $\ell$ yields,
\be
  \ell {dm^2\over d\ell}\Big|_{m\ell\gg1} = 
  m^2\(-{(\epsilon+2)(\epsilon-4)(\epsilon-2)(6(1-\xi)+\epsilon(3\xi-2))\over 24 \,(m\ell)^2} + \mathcal{O}\big((m\ell)^{-4}\big)  \) ~,
  \label{GdmdL}
\ee
Substituting this expression and \reef{tstar} into \reef{main2} gives
\bea
 \ell {dS_\mt{EE}\over d\ell}\Big|_{m\ell\gg1}&=&{(4\pi)^{{\epsilon\over2}-1} (\epsilon-2)\big(6(1-\xi)+\epsilon(3\xi-2)\big) \over 12\,\epsilon\,(3-\epsilon)} 
\, \Gamma\(2+{\epsilon\over2}\) \,N A_\Sigma m^{2-\epsilon} \Big( 1 + \mathcal{O}\big((m\ell)^{-2}\big) \Big)
 \non
 &+&4\pi(\epsilon -2)\,N\, A_\Sigma \bar\kappa  +s_0 ~,
     \labell{WFent}
\eea

The first thing to note about \reef{WFent} is that the `area law' term does not have any $\mu$-dependance, and therefore it is not sensitive to the constant of proportionality in the relation $\mu\sim\ell^{-1}$, \ie as expected, the value of entanglement entropy at the fixed point is invariant under reparametrizations of the RG trajectory.

It is instructive to compare \reef{WFent} with its counterpart in \cite{Sachdev}. The results in  \cite{Sachdev} are intrinsic to the Wilson-Fisher fixed point since their setup, unlike ours, confines the RG flow to the IR end. The geometry in \cite{Sachdev} is flat, and therefore the gravitational coupling, $\bar\kappa$, which appears in \reef{WFent} is absent. In addition, as we argue  in section \ref{sec:bndry}, the computation of \cite{Sachdev} corresponds to $\xi=1$ at the Wilson-Fisher fixed point. Thus, to leading order in $\epsilon\ll 1$, Eq.~\reef{WFent} reduces to
\be
  \ell\frac{d S_\mt{EE}}{d \ell}\Big|_{m\ell\gg 1} \simeq -N\,{A_\Sigma m_\mt{$\infty$}^{2-\epsilon}\over 72\pi}  \quad\Leftrightarrow \quad  
  S_\mt{EE} \simeq -N\,{A_\Sigma m_\mt{$\infty$}^{2-\epsilon}\over 144\pi}  ~,
  \labell{MFS}
\ee
where $m_\mt{$\infty$}$ is the mass gap in the limit of flat space. Eq. \reef{MFS} is in agreement with \cite{Sachdev}. A simple derivation of (\ref{MFS}) was later given by Casini, Mazzitelli and Test\'e \cite{Casini:2015aa}. The authors noticed that to leading order in $\epsilon$, the anomalous dimension vanishes at the Wilson-Fisher fixed point, and thus (\ref{MFS}) may be found from the entanglement entropy for a free field. One distinction between our work and that of \cite{Casini:2015aa}, is that \cite{Casini:2015aa} advocates  that the entanglement Hamiltonian has a discontinuous jump at the UV fixed point. Namely, that the entanglement entropy takes the value for the minimally coupled scalar at the free field endpoint of the RG trajectory, whereas it takes the conformally coupled value at all other locations. In our setup, the entanglement entropy behaves smoothly along the entire RG trajectory. As found from the beta functions (\ref{beta}), starting with either a minimally or nonminimally coupled field in the UV leads to the conformally coupled field in the IR. 

Let us now expand the numerical coefficient of the `area law' term in \reef{WFent} in $\epsilon\ll 1$
\bea \label{S4d}
 \ell {dS_\mt{EE}\over d\ell}\Big|_{m\ell\gg1}&=&{N\, A_\Sigma m^{2-\epsilon}\over 12} 
 \( {-1\over 6\pi}+(\xi-1)\({1\over \pi\epsilon} - {2\gamma+2/3-2\log(4\pi)\over 4\pi}\) +\mathcal{O}\big(\epsilon, (m\ell)^{-2}\big) \)  
  \non
  &+&4\pi(\epsilon -2)\,N\, A_\Sigma \bar\kappa + s_0 ~.
\eea
In the next section, we will see that the $1/\epsilon$ term in (\ref{S4d}) is associated with UV degrees of freedom. \footnote{The IR theory is empty, as we have only massive degrees of freedom which decouple in the deep IR.}  The presence of this  UV remnant is a result of using the full energy-momentum tensor \reef{trace2} to calculate entanglement entropy at any scale $\mu\sim\ell^{-1}$. 

To isolate entanglement entropy intrinsic to the scale $\ell$, one needs to use some kind of subtraction scheme. There is no unique or preferred choice of such a scheme. Renormalized entanglement entropy \cite{Liu:2012eea} is one possibility. This prescription proved to be particularly powerful in three dimensions, and was used in the proof \cite{Casini:2012ei} of the F-theorem \cite{Jafferis:2011zi}, see also \cite{Casini:2015woa}. Unfortunately, it is not clear that renormalized entanglement entropy has analogous properties, such as monotonicity, in integer dimensions higher than the $3$; nor is it clear how to apply it in non-integer dimensions.

For the particular choice of $\xi=1$, the theory is conformally coupled along the entire RG trajectory, and the contribution of UV degrees of freedom to the `area law` in the vicinity of $d=4 \, (\epsilon=0)$ vanishes. Since in this case \reef{WFent} is not contaminated by UV physics,  it can be used to find an approximation for entanglement entropy at the interacting IR fixed point in three dimensional flat space. Substituting $\xi=\epsilon=1$ and $\bar\kappa=0$ into \reef{WFent}, gives
\bea
 S^\mt{IR}_\mt{EE}\Big|_{d=3}&\simeq& -{N\over 64}  m  A_\Sigma   ~.
\eea

We note that the constant  $\bar\kappa$ is arbitrary; it can be exchanged for the entanglement entropy at some scale $l_1$. 
A choice that would seem natural is to demand that entanglement entropy vanishes in the deep IR (as a result of the mass gap, all degrees of freedom decouple in the IR, and the Wilson-Fisher fixed point is thus empty).  This  results in
\be
 S_\mt{EE}\Big|_{m\ell\gg 1}=0 \quad \Rightarrow \quad \bar\kappa
 =m^{2-\epsilon}{(4\pi)^{{\epsilon\over2}-2} \big(6(\xi-1)-\epsilon(3\xi-2)\big) \over 12\,\epsilon\,(3-\epsilon)} 
\, \Gamma\(2+{\epsilon\over2}\)~.
 \labell{kvalue}
\ee
The above subtraction scheme is special to a curved manifold with nondynamical gravity, where there is an extra parameter $\bar\kappa$. However in flat space  $\bar\kappa=0$, and  one is  forced to adopt a different subtraction scheme. Another drawback of the choice (\ref{kvalue}) is that it modifies entanglement entropy at all points along the RG trajectory, and not only in the deep IR limit. The latter makes it difficult to extrapolate the results for an `area law' on a sphere to flat space which has no analog of $\bar\kappa$. In what follows we simply leave $\bar{\kappa}$ unspecified.

\subsection{Gaussian fixed point} \label{sec:Gaussian}
This time we expand \reef{phi2} in $m\ell\ll \epsilon \ll 1$. In this regime the theory flows to the Gaussian UV fixed point where  $u$ asymptotically vanishes, $u\ll \epsilon$. From the gap equation, 
\bea
   \langle \phi^2 \rangle\Big|_{m\ell\ll1}&=&
   \al_1 m^2 \ell^{\epsilon} 
   \( {1\over (m\ell)^2} + \al_2
   + \mathcal{O} \( (m\ell)^2 \) \)
   ~,
  \labell{phi2IR}
\eea
where $\lambda_0=\lambda|_{m\ell=0}$ and for brevity we introduced the following constants
\bea
 \al_1&\equiv& { \Gamma\big( {\epsilon -2 \over 2} \big)\Gamma(\lambda_0) \Gamma(\bar\lambda_0) \cosh\big( {\pi(\lambda_0-\bar\lambda_0) \over 4 i} \big)\over 
   \pi (4\pi)^{4- \epsilon\over 2} }  ~,
   \non
 \al_2&\equiv& 2 \,{\psi(\bar\lambda_0) - \psi(\lambda_0) + i \pi \tanh\big( {\pi(\lambda_0-\bar\lambda_0) \over 4 i} \big) \over \lambda_0-\bar \lambda_0}  ~,
\eea
where $\psi(\lambda)$ is the digamma function. The two terms that we kept in \reef{phi2IR} are the only ones that diverge as $\epsilon\to 0$. We now substitute this expansion into \reef{gapeq} and use \reef{renorm}
\be
 m^2\( {1\over u} - {1\over 8\pi^2\epsilon} \)
 ={t\over u} + \alpha_1 m^2 (\mu\ell)^\epsilon \( {1\over (m\ell)^2} + \al_2 +\mathcal{O}((m\ell)^2) \) + \({\eta\over u} - {1-\xi\over 48\pi^2 \epsilon} \) R~.
 \labell{gaussgap}
\ee
Solving the RG equation for $\eta$ and $u$ gives
\be
 {\eta\over u}={1-\xi\over 48\pi^2\epsilon}  ~.
 \labell{etau}
\ee
Hence, we can drop the last term in \reef{gaussgap},
\be
 \ell {d m^2\over d\ell}\Big|_{m\ell\ll1} = u \, \al_1 m^2 (\mu\ell)^\epsilon \( {\epsilon-2 \over (m\ell)^2} + \epsilon\,\al_2 + \mathcal{O}\big((m\ell)^2\big)  \) + \mathcal{O}(u^2) ~.
  \labell{WFdmdL}
\ee
Substituting \reef{gaussgap} and \reef{WFdmdL} into \reef{main2}, we finally deduce,
\be
 \ell {dS_\mt{EE}\over d\ell}\Big|_{m\ell\ll1} = - {2\pi \alpha_1(2-\epsilon) \over (4-\epsilon)(3-\epsilon)} 
 \,NA_\Sigma m^{2} \ell^\epsilon \Big(1+\mathcal{O}(u) \Big)
 +4\pi(\epsilon -2)\,N\, A_\Sigma \bar\kappa +\mathcal{O}\big((m\ell)^4\big) +s_0~.
\ee
Expanding the coefficient of $A_\Sigma m^{2} \ell^\epsilon$ in $\epsilon\ll 1$, we get
\be
 \ell {dS_\mt{EE}\over d\ell}\Big|_{m\ell\ll1} = N\,A_\Sigma m^2\ell^\epsilon \({\xi-1\over 12 \pi\epsilon} + \mathcal{O}(\epsilon, u) \)
  +4\pi(\epsilon -2)\,N\, A_\Sigma \bar\kappa +\mathcal{O}\big((m\ell)^4\big) +s_0 ~.
 \label{EEgauss}
\ee

As in the previous case, the $\mu$ dependence drops out of the final answer. While we implicitly assumed that $\mu\sim\ell^{-1}$, the final answer for entanglement entropy at the fixed point should not be sensitive to the $\mathcal{O}(1)$ coefficient of proportionality between $\mu$ and $\ell$. Furthermore, the $1/\epsilon$ term is the same term that  appears in \reef{WFent}, which, as mentioned in the previous section and now seen explicitly  in \reef{EEgauss},  represents entanglement entropy of the UV degrees of freedom. 

For a minimally coupled scalar field, $\xi=0$,  and we recover the well-known universal `area law', 
in agreement with \cite{Sachdev} and \cite{Hertzberg:2010uv,Huerta:2011qi,Lewkowycz:2012qr} (this is just $N$ times the answer for a free scalar). If, however, the field is non-minimally coupled, we get a different answer which vanishes at the conformal point $\xi=1$. In Sec.~\ref{sec:FMS2}, we show how to generalize the calculation at the Gaussian fixed point  presented in \cite{Sachdev}, so as to take into account  the contribution from non-minimal coupling.

\subsection{Along the RG trajectory} \label{sec:RG}
In this section we write down the entanglement entropy for the $O(N)$ model for a conformally coupled scalar, at leading order in $1/N$ in dimension $4-\epsilon$, for any sphere of radius $\ell$. The ingredients have been worked out in the previous sections; here we just collect them. 

Solving the RG equation (\ref{beta}) gives
\be
u(\mu) = \frac{u^*}{1+ \(\frac{\mu}{\mu_0}\)^{\epsilon}}~,
\ee
where $\mu_0$ is an arbitrary constant scale and $u^*$ is $u$ at the Wilson-Fisher fixed point. We want to take $\mu = 1/\ell$, and we let $\mu_0 = \ell_0^{-1}$. \footnote{Since the only scale is $\ell$, it must be that $\mu$ is proportional to $\ell^{-1}$. The constant of proportionality can be absorbed into $\mu_0$.} We will write this as
\be \label{usol}
u(\ell) = \frac{u^*}{1+ \(\frac{\ell_0}{\ell}\)^{\epsilon}}~.
\ee
Note that the entanglement entropy will contain the constant $\ell_0$. This is analogous to how correlation functions contain an arbitrary scale which is calibrated through some measurement. In our context, it means entanglement entropy needs to me measured at one value of $\ell$, and can then be predicted at all other values. 

The gap equation is given by
\be \label{gapreg}
m^2\(\frac{1}{u} - {1\over u^*} \) = \frac{t}{u} + \ell^{-\epsilon} \langle \phi^2\rangle ~,
\ee
where for simplicity we chose $\xi=1$ (hence, $\eta=0$ along the entire RG flow) and $\langle \phi^2 \rangle$ is given by (\ref{phi2}). Also, 
\be \label{dphi}
\ell \frac{d}{d\ell} \big( \ell^{-\epsilon}\langle \phi^2\rangle\big) = \(\ell \frac{d \lambda}{d\ell} \frac{\del }{\del \lambda} - 2\) \big( \ell^{-\epsilon}\langle \phi^2\rangle\big) ~,
\ee
where  $\lambda$ is a function of $m\ell$ and is given by (\ref{eq:lambda}). Differentiating (\ref{gapreg}) and using (\ref{dphi}), we can solve for $\ell \frac{d m^2}{d \ell}$. 

The entanglement entropy is given by (\ref{main2}), where we substitute $u(\ell)$ given by (\ref{usol}) and $t/u$ given by (\ref{gapreg}). We thus have a complete expression for entanglement entropy in terms of the mass $m$ and radius $\ell$. 

\section{Boundary perturbations}
\label{sec:bndry}
In the previous Section we computed  entanglement entropy along the entire RG flow, and in particular in the proximity of the fixed points. The entanglement entropy was found to be sensitive to the non-minimal coupling parameter $\xi$. This sensitivity is  robust: it persists in the flat space limit, and away from the UV fixed point.

In light of these results, in this Section we revisit the replica-trick calculations of entanglement entropy near fixed points. The fact that entanglement entropy depends on $\xi$, even in flat space, is  manifest in the context of the replica-trick, and is a consequence of the curvature associated with the conical singularity. What is unclear is if this boundary term which gives $\xi$ dependance is real, or an artifact of the replica-trick which should be discarded. The results of Sec.~\ref{sec:EE} suggest the former. 

In Sec.~\ref{sec:heat} we review the replica-trick calculation of entanglement entropy for a free scalar field using heat kernel techniques. In Sec. \ref{sec:FMS} we review a calculation of Metlistski, Fuertes, and Sachdev \cite{Sachdev} which finds that loop corrections generate a boundary term, and we argue that this has a simple interpretation as  the classical boundary term of Sec.~\ref{sec:heat}  due to the curvature of the conical singularity. In Sec.~\ref{sec:FMS2} we generalize the calculation of \cite{Sachdev} of entanglement entropy at the Gaussian fixed point, so as to incorporate non-minimal coupling, and find agreement with the results of Sec. \ref{sec:EE}. 

\subsection{Replica trick: Free energy} \label{sec:heat}
Here we preform the standard replica trick calculation for a free massive scalar \cite{Kabat95, Callan:1994py, SusUgl94}. Recall that entanglement entropy is computed from the variation of the free energy, 
\be \label{Sfree}
S= \(\beta \frac{\partial}{\partial \beta} - 1\)\Big|_{\beta = 2\pi} (\beta F)~,
\ee
where the effective action $(\beta F)$ is evaluated on a space which is a cone with a deficit angle $2\pi - \beta$.  Eq.~\reef{Sfree} is just the standard thermodynamic equation for entropy; the need to vary the temperature away from $1/2\pi$ introduces a conical singularity at the origin. The effective action, after integrating out the matter, is expanded in derivatives of the metric, 
\be
\beta F = -\frac{1}{2}\int_{\mathcal{M}} \int_{\delta^2}^{\infty} \frac{ds}{(4\pi s)^{d/2}} e^{-s m^2} \(\frac{c_0}{s} + c_1 R + \mathcal{O}(s)\)
\ee
The relevant term is the one proportional to the scalar curvature, whose integral on a cone is $\int_{\mathcal{M}} R = 2\mathcal{A}_{\Sigma}\, (2\pi - \beta)$.
Thus the entropy is, 
\be \label{eq:Skabat}
S = 2 \pi\, c_1\, \mathcal{A}_{\Sigma}\, \int_{\delta^2}^{\infty} \frac{ds}{(4\pi s)^{d/2}} e^{-s m^2}~.
\ee
Specializing to $d=4$ and expanding the exponential in (\ref{eq:Skabat}) to extract the $\log$ divergent piece, we get,
\be \label{eq:Skabat2}
S = \frac{c_1}{4 \pi}\, m^2\, \mathcal{A}_{\Sigma}\, \log(\delta)~.
\ee
For the minimally coupled scalar $c_1 =1/6$, while for the conformally coupled scalar $c_1=0$ \cite{Vas}.  

This computation is, of course, not new. However, it conflicts with the belief that (in the flat space limit) entanglement entropy should, like correlation functions, be unable to distinguish a minimally from nonminally coupled scalar field. The agreement of (\ref{eq:Skabat2}) with the independent results of Sec.~\ref{sec:EE} suggests Eq.~\reef{eq:Skabat2} should be taken seriously. 

\subsection{Loops generate a boundary term} \label{sec:FMS}
In \cite{Sachdev} entanglement entropy is computed using the replica-trick approach. Introducing the standard replica symmetry around a given codimension-two entangling surface, the entanglement entropy is given by
\be \label{eq:Sreplica}
S_\mt{EE} = \lim_{n \rightarrow 1} \frac{1}{1-n}\log{\frac{Z_n}{Z^n_1}},
\ee
where $Z_n$ is the partition function of the theory on an $n$-sheeted Riemanian manifold, $\mathcal{M}_n$. The entangling surface is where the $n$ sheets are glued together, and is the location of the conical singularity. In computing correlation functions on $\mathcal{M}_n$, it is important to note that $\mathcal{M}_n$ has separate translation symmetries in the directions parallel to the entangling surface, and in the directions orthogonal to it. Clearly, the entangling surface is a special location. 

In \cite{Sachdev} the authors consider the loop corrections that a correlation function on $\mathcal{M}_n$, such as a two-point function, will receive from interactions. They find that as a result of loops in the vicinity of the entangling surface, new divergences are generated, forcing the introduction of a boundary counter-term in the action, 
\be
 \delta I = {c\over 2} \int_\Sigma \vec{\phi}^{\,2}~,
 \labell{ind}
\ee
where the integral is restricted to the entangling surface, $\Sigma$. Performing an RG analysis  gives (to leading order in the large-$N$ expansion) the renormalized coupling, $c$, at the Wilson-Fisher fixed point  \cite{Sachdev}
\be
 c^*=-{2\pi\over 3} (n-1)~.
\ee

In fact, as we now show, this result has a simple interpretation in terms of the conformal coupling to the background metric. Recall that the action (\ref{action}) contains the term $\int {1\over 2}(\xi\eta_c+\eta_0) R \phi^{\,2}$. Solving the RG equations, we found in Sec.~\ref{sec:beta} that at the Wilson-Fisher fixed point $\eta^* + \xi \eta_c =1/6$ to leading order in $\epsilon$. At the Wilson-Fisher fixed point this part of the action is therefore, 
\be \label{eq:deltaI}
\delta I = \frac{1}{12} \int R\ \phi^2~.
\ee
As we have mentioned before, and is reflected in (\ref{eq:deltaI}), the RG flow leads to conformal coupling in the IR, regardless of the non-minimal coupling $\xi$ in the UV. Now we need to evaluate (\ref{eq:deltaI}) on the background $\mathcal{M}_n$. Recall that to linear order in $(n-1)$, the expansion of the curvature scalar, $R^{(n)}$, on a replicated manifold is given by \cite{Fursaev:1995ef}
\be
 R^{(n)}=R^{\mt{reg}} + 4\pi (1-n) \delta_\Sigma +\ldots~,
 \labell{curv}
\ee
where $R^{\mt{reg}}$ is the regular curvature  in the absence of the conical defect. The term $\delta_\Sigma$ is a two-dimensional delta function with support on the entangling surface $\Sigma$, and reduces the $d$-dimensional integral over the entire manifold to an integral over the entangling surface.~\footnote{The higher order terms in \reef{curv} are ambiguous \cite{Fursaev:1995ef}, and therefore in general only linear order terms in $(n-1)$ are reliable.}
Inserting \reef{curv} into (\ref{eq:deltaI}) gives
\be
 \delta I= - {\pi\over 3}(n-1) \int_\Sigma \phi^2 ~,
\ee 
in agreement with (\ref{ind})  \ie the induced boundary perturbation \reef{ind} is just the conformally coupled action evaluated on the conical defect.

\subsection{Including non-minimal coupling} \label{sec:FMS2}
In Sec.~\ref{sec:heat} we showed that  entanglement entropy  for a free scalar is different depending on if the scalar is minimally or conformally coupled. The calculation was done using the replica trick, combined with heat kernel techniques. In  \cite{Sachdev}, entanglement entropy was calculated at the Gaussian fixed point using the replica trick, and by evaluating a 2-pt function on the conical background. The computation in  \cite{Sachdev} was implicitly for a minimally coupled scalar. For completeness, here we generalize the calculation to incorporate non-minimal coupling,

At the Gaussian fixed point the action for the $O(N)$  model simplifies to
\be
I = \int \left(\frac{1}{2}\big(\partial \vec\phi \, \big)^2 + \frac{t}{2}\vec{\phi}^{\,2} + \frac{1}{2}\,\xi\,\eta_cR\vec\phi^{\,2}\right). 
\ee
It follows from (\ref{eq:Sreplica}) that,
\be
 { d \over dt} S_\mt{EE} = \lim_{n \rightarrow 1} \frac{1}{1-n}  \frac{d}{dt}\log{\frac{Z_n}{Z^n_1}}. 
\ee
Using \reef{curv} to expand $Z_n$, and keeping only terms of order $n-1$, 
\be
Z_n =\int \mathcal{D}\phi \, 
\(1 + 2\pi\eta_c (n-1)\int_\Sigma \phi^2 + \ldots \)
 \exp\(-\int_{\mathcal{M}_n} \frac{1}{2}(\partial \vec\phi \, )^2 + \frac{t}{2}\vec{\phi}^{\,2} + \frac{1}{2}\,\xi\,\eta_cR^\mt{reg}\vec\phi^{\,2}\)~.
\ee
Now taking a derivative with  respect to $t$ gives
\be \label{eq:ddt}
\frac{d}{dt} \log Z_n= -\frac{1}{2}\int_{\mathcal{M}_n} \langle \vec\phi^{\,2} \rangle_n  -  \pi \xi\eta_c (n-1) \int_{\mathcal{M}_1}  \int_\Sigma  \langle \, \vec\phi^{\,2} \, \vec\phi^{\,2} \, \rangle_1 + \ldots 
\ee
Note that to leading order in $(n-1)$, we can take both the integral and the two-point function in Eq.~\reef{eq:ddt} to be over a single sheet. Thus, we obtain 
\bea
{ dS_\mt{EE}\over dt} =\lim_{n\rightarrow 1}\frac{1}{n-1}\bigg[ \frac{1}{2}\(\int_{\mathcal{M}_n} \langle \vec\phi^{\,2} \rangle_n 
- n \int_{\mathcal{M}_1} \langle \vec\phi^{\,2} \rangle_1  \)  +  \pi\xi \eta_c (n-1) \int_{\mathcal{M}_1}  \int_\Sigma  \langle \, \vec\phi^{\,2} \, \vec\phi^{\,2} \, \rangle_1 \bigg].
\labell{dSdt}
\eea

It is convenient to specialize to the case of a planar entangling surface embedded in flat space; the  `area law' terms are insensitive to this choice. For $\xi = 0$, only the first two terms in \reef{dSdt} survive and evaluate to \cite{Sachdev}
\be
 \lim_{n\rightarrow 1}\frac{1}{n-1}\bigg[ \frac{1}{2}\(\int_{\mathbb{R}_n^d} \langle \vec\phi^{\,2} \rangle_n 
- n \int_{\mathbb{R}^d} \langle \vec\phi^{\,2} \rangle_1  \) \bigg] = \frac{-N}{24\pi \epsilon} ~,
\ee
where the $1/\epsilon$ pole signals that there is a logarithmic divergence as $\epsilon\to 0$.

For $\xi\neq 0$, the last term in (\ref{dSdt}) must included. To evaluate it, we first note that  
\bea
 &&\langle \phi_a(x) \phi_c(0)\rangle_1 = \frac{\delta_{ac}}{(2\pi)^{d\over2}} \({\sqrt{t}\over |x|}\)^{d-2\over 2} K_{d-2\over 2}\big(\sqrt{t\, x^2}\big) ~, 
  \non
 &&\langle \vec\phi^{\,2}(x) \vec\phi^{\,2}(0)\rangle_1 =  \frac{2\, N}{(2\pi)^d} \({t\over x^2}\)^{d-2\over 2} K_{d-2\over 2}^2\big(\sqrt{t\, x^2}\big)~.
\eea
where $K_{d-2\over 2}$ is the modified Bessel function. Hence,
\be
 \pi\xi\eta_c \int_{\mathcal{M}_1} \int_\Sigma   \langle \, \vec\phi^{\,2} \, \vec\phi^{\,2} \, \rangle_1 = {N \xi\,\eta_c \over (4\pi)^{d-2\over 2} \Gamma\({d\over 2}\)} 
 \, t^{d-4\over 2} A_\Sigma
 \int_0^\infty dr \, r \,K_{d-2\over 2}^2(r) 
 = {N \xi\,\eta_c \Gamma\({4-d\over 2}\)\over 2 (4\pi)^{d-2\over 2} } 
 \, t^{d-4\over 2} A_\Sigma ~.
 \labell{phi2phi2}
\ee
where we introduced a dimensionless variable $r=\sqrt{t\, x^2}$. Substituting $d=4-\epsilon$ and expanding in $\epsilon\ll 1$ gives
\be
 \pi\xi\eta_c \int_{\mathcal{M}_1} \int_\Sigma   \langle \, \vec\phi^{\,2} \, \vec\phi^{\,2} \, \rangle_1 
 = {N \xi \over 24\pi \epsilon} 
 \, A_\Sigma +\mathcal{O}(\epsilon^0) ~.
\ee
Combining the above results, we recover the $1/\epsilon$ term in \reef{EEgauss}. Note also that the integrand on the right hand side of \reef{phi2phi2} decays exponentially fast in the IR. In particular, the $1/\epsilon$ enhancement comes entirely from the UV regime ($r\sim 0$).

\subsubsection*{Comments}
We close with a few comments. The question of if a minimally and non-minimally coupled scalar have the same entanglement entropy in flat space is an old one, see \eg \cite{Larsen:1995ax}, in which interest has recently revived 
\cite{Klebanov:2012va}, \cite{Nishioka:2014kpa, Lee:2014zaa, Herzog:2014fra, Dowker:2014zwa, MRS, Rosenhaus:2014ula, Casini:2015aa,Ben-Ami:2015aa}. A practical question concerns the computation of entanglement entropy on a lattice
\cite{Srednicki:1993im,Huerta:2011qi, Lohmayer:2009sq} and whether certain boundary terms should be included even for a scalar theory. Such boundary terms were advocated in \cite{Donnelly:2011hn,Casini:2013rba,Donnelly:2014gva,Donnelly:2014fua} for gauge theories. The lattice calculation is carried out in flat space and naively the non-minimal coupling plays no role. To what extent this claim is true requires further investigation. In particular, it is essential to understand how one splits the Hilbert space in a conformally invariant way. We note that for a CFT the lattice computation of the universal entanglement entropy whose coefficient is fixed by the trace anomaly is not affected by this issue. In particular, for a massless free scalar field it seems not to depend on whether one uses the canonical or the improved energy-momentum tensor.

In Sec.~\ref{sec:heat} we found a term localized on the tip of the cone, originating from the non-minimal coupling to the background geometry. In  \eg \cite{Solodukhin:1996jt,Hotta:1996cq,Solodukhin:2011gn}, the authors also found such  a contribution, \footnote{ Such a contribution was discussed in these works in the context  of  the leading non universal ($A/\delta^2$) area law piece of entanglement entropy. However,  the term in question contributes to the universal part of entanglement entropy as well. } but they regarded it as an artifact of the replica-trick and discarded it. However, our result \reef{EEgauss} relies on neither the replica trick nor free field calculations, and is consistent with the presence, but not the absence, of the term on the tip of the cone. 
Finally, the analysis  of \cite{Sachdev} in the interacting case did not include the contribution of the non-minimal coupling at the tip of the cone (or alternatively, their scalar field is implicitly minimally coupled). Yet, their results imply that  quantum fluctuations on the conical background force the introduction of the boundary counter terms which, as we have argued, have a simple interpretation in terms of induced non-minimal coupling to the background geometry. And, as the RG equations show, this occurs even if the theory is minimally coupled in the UV. We conclude that QFT on the cone background is incomplete without the inclusion of boundary terms.

\acknowledgments  
We thank O. Aharony, H.~Casini, S.~Giombi, I.~Klebanov, J.~Koeller, Z.~Komargodski, M.~Metlitski, R.C.~Myers, and H.~Neuberger for helpful discussions and correspondence. 
The work of VR is supported by NSF Grants PHY11-25915 and PHY13-16748.
The work of MS is supported by NSF Grant PHY-1521446 and the Berkeley Center for Theoretical Physics. 
The work of SY is supported in part by the ISF center of excellence program (grant 1989/14), BSF (grant 2012/383) and GIF (grant I-244-303.7-2013)

\appendix

\section{$F_\Lambda$ counterterm}
\label{Flam}

In this Appendix we carry out an independent calculation of  $F_\Lambda$ in order to verify \reef{FLamFk}. As argued in \cite{Brown:1980qq}, $F_\Lambda$ is directly related to the divergences of the massless correlator $\langle [\vec\phi^2](x) [\vec\phi^2](y) \rangle$, where $[\ldots]$ denotes a renormalized composite operator. Such divergences are removed by adding a counter term proportional to a delta function, 
\be
{1\over 4} \langle [\vec\phi^2](x) [\vec\phi^2](y) \rangle + \mu^{-\epsilon} N\,A(u,\epsilon) \, \delta^d(x,y) 
= {1\over 4} \({t_0\over t }\)^2 \langle \vec\phi^2(x) \vec\phi^2(y) \rangle + \mu^{-\epsilon} N\,A(u,\epsilon) \, \delta^d(x,y)
=\text{finite} ~,
\ee
where $A(u,\epsilon)$ has poles in $\epsilon$, and \cite{Brown:1980qq}\footnote{Since we are using Euclidean signature there is a difference in the relative sign between $F_\Lambda$ and $A$ in comparison to \cite{Brown:1980qq}.}
\be
  F_\Lambda(u,\epsilon)=-{1\over 2} \({t \over t_0}\)^2 A(u,\epsilon)~.
\ee

\begin{figure}
\begin{center}
\includegraphics[width=10cm,height=3cm]{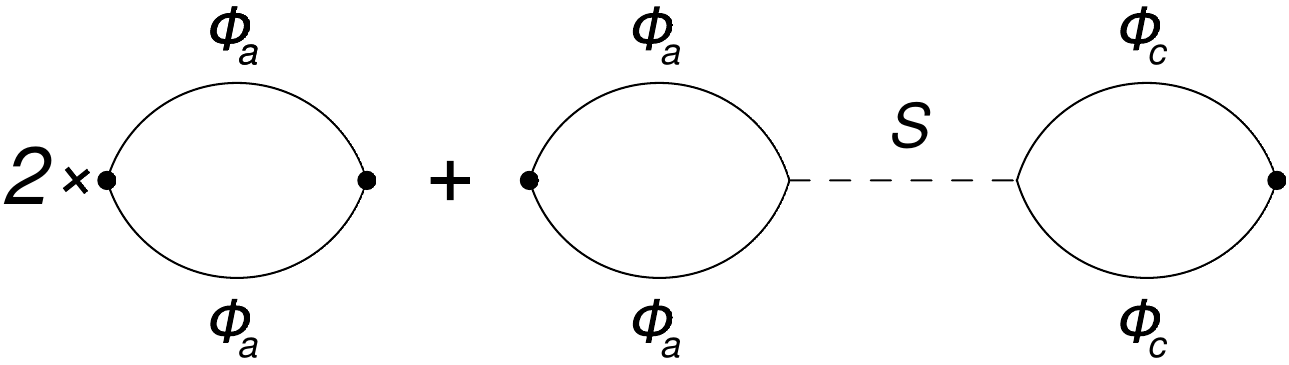}
\caption{Two diagrams of order $N$ which contribute to the Green's function $\langle \vec\phi^2(x) \vec\phi^2(y) \rangle$. Solid and dashed lines represent propagators of the scalar, $\vec \phi$, and auxiliary field, $s$, respectively. The dots represent insertions of $\vec\phi^2$.}
\label{diag}
\end{center}
\end{figure}

To leading order in  large-$N$, there are two diagrams that contribute to $A(u,\epsilon)$, see Fig. \ref{diag}. 
To evaluate these diagrams we have to calculate the propagator of the auxiliary field $s$. To this end, we expand the generating functional \reef{part} around the saddle point $s=\bar s + s'$ and use cyclicity of $\text{Tr}$, \eg
\be
 \text{Tr} \ln \hat{O}_{s}=\text{Tr} \ln \(\hat{O}_{\bar s}(1 + \hat{O}_{\bar s}^{-1}s')\) = \text{Tr} \ln \hat{O}_{\bar s} +\text{Tr}  \big(\hat{O}_{\bar s}^{-1}s'\big) 
 -{1\over 2}\text{Tr}  \big(\hat{O}_{\bar s}^{-1}s' \hat{O}_{\bar s}^{-1} s'\big) +\ldots
\ee
Constant terms in the expansion of the action are part of the normalization and can therefore  be suppressed. Linear terms vanish since we are expanding around the saddle point. As a result, the expansion of the action to second order is given by,
\be
 S_{\text{eff}}(s')=
 -{N\over 4}\int d^dx\sqrt{g(x)}\int d^dy \sqrt{g(y)} \, s'(x)\( D^2(x,y) + {\delta^d(x,y)\over u_0}\)s'(y) + \ldots ,
 \labell{quadr}
\ee 
where $D(x,y)\equiv\langle x|\hat{O}_{\bar s}^{-1}|y\rangle$ denotes the {\it full} propagator of the scalar field, $\vec\phi$, to leading order in the large-$N$ expansion. 

The propagator of the auxiliary field $s'$ can be easily obtained by inverting the above quadratic form. On a sphere, such an inversion can be done in closed form by noting that this quadratic form is diagonal in the basis of spherical harmonics. We will not carry out the full calculation since we only need the $\delta$-functions in the propagator of $s'$,  for which a short distance expansion and flat space approximation are sufficient. In particular,
\be
 D^2(x-y)= {\mu^{-\epsilon}\over 8\pi^2\epsilon} \, \delta^d(x-y)+ \mathcal{O}(\epsilon^0) ,
 \labell{bubble}
\ee
where we suppressed (finite) terms without $\delta$-functions. Substituting into \reef{quadr}, and using \reef{renorm}, yields
\be
 \langle s'(x) s'(y) \rangle = -{2 u  \mu^{\epsilon}  \over N}  \delta^d(x-y) + \ldots 
 \labell{sprop}
\ee
As expected, \reef{renorm} renders the effective action finite to leading order in  large-$N$.
Using now \reef{bubble} and \reef{sprop}, the diagrams in Fig. \ref{diag} can be readily evaluated. The final result reads
\be
 A=-{1\over (4\pi)^2 \epsilon} \({t_0\over t }\)^2\(1-{u \over 8\pi^2\epsilon}\) \quad \Rightarrow \quad 
 F_\Lambda(u,\epsilon)={1\over 2(4\pi)^2 \epsilon} \(1-{u \over 8\pi^2\epsilon}\) ~,
 \labell{FLam}
\ee
in full agreement with \reef{FLamFk} and \cite{Fei:2015oha}.

\bibliographystyle{utcaps}

\bibliography{lib}

\end{document}